\documentclass[preprint]{elsarticle}
\usepackage{setspace}

\usepackage{amssymb}
\usepackage{graphicx}% Include figure files
\usepackage[outdir=./]{epstopdf}
\usepackage{dcolumn}% Align table columns on decimal point
\usepackage{amsmath}
\usepackage{epsfig}
\usepackage{array}
\usepackage{epstopdf}
\usepackage{booktabs,hyperref}
\hypersetup{colorlinks=true}
\usepackage{multirow}
\usepackage{anyfontsize}
\usepackage{caption,setspace}
\usepackage{subcaption}
\usepackage{longtable}
\usepackage{color,soul}

\usepackage{mwe}
\usepackage{pifont} 
\usepackage[table]{xcolor}
\usepackage{enumitem}  
\usepackage[a4paper,bindingoffset=0.2in,%
            left=1.25in,right=1.25in,top=1in,bottom=1in,%
            footskip=.25in]{geometry}
\usepackage{units}

\providecommand{\numberTblEq}[1]{\refstepcounter{tblEqCounter}\label{#1}\thetag{\thetblEqCounter}}

\biboptions{numbers,sort&compress}

\definecolor{gray(x11gray)}{rgb}{0.75, 0.75, 0.75}

\begin{document}



\begin{frontmatter}
\title{Development of a Comprehensive Physics-Based Battery Model and Its Multidimensional Comparison with an Equivalent-Circuit Model: Accuracy, Complexity, and Real-World Performance under Varying Conditions}
%\title{Multidimensional Comparison of Physics-Based and Equivalent-Circuit Models for Lithium-Ion Batteries: Accuracy, Complexity, and Real-World Performance Across Varying Conditions}

\author[SJTU,NEL]{Guodong Fan\corref{cor}}
\ead{guodong.fan@sjtu.edu.cn}
\author[SJTU,NEL]{Boru Zhou}
\author[SJTU,NEL]{Chengwen Meng}
\author[SJTU,NEL]{Tengwei Pang}
\author[SJTU,NEL]{Xi Zhang}
\author[CATL]{Mingshu Du}
\author[CATL]{Wei Zhao}

\cortext[cor]{Corresponding author}
\address[SJTU]{School of Mechanical Engineering, Shanghai Jiao Tong University, 800 Dongchuan Road, Shanghai, 200240, China}
\address[NEL]{National Engineering Research Center of Automotive Power and Intelligent Control, Shanghai Jiao Tong University, Shanghai, 200240, China}
\address[CATL]{Battery Management System Department, Contemporary Amperex Technology Ltd. (CATL), Ningde, Fujian 352000, China}

\doublespacing

\begin{abstract}

This paper develops a comprehensive physics-based model (PBM) that spans a wide operational range, including varying temperatures, charge/discharge conditions, and real-world field data cycles. The PBM incorporates key factors such as hysteresis effects, concentration-dependent diffusivity, and the Arrhenius law to provide a realistic depiction of battery behavior. Additionally, the paper presents an in-depth analysis comparing the PBM with an equivalent-circuit model (ECM) for accurately capturing the dynamics of lithium-ion batteries under diverse operating conditions. To ensure a fair comparison, both the PBM and ECM are rigorously calibrated and validated through parameter identification and testing across 55 different operating conditions. To the best of the authors' knowledge, this represents the most comprehensive model calibration and validation effort for PBM and ECM in the literature to date, encompassing large temperature variations (-20 to 40°C), various charging/discharging C-rates, and real-world driving cycles. Comparative analysis between the PBM and ECM highlights key differences in accuracy, computational complexity, parameterization requirements, and performance under varying temperature conditions. appropriate models for battery management applications.

%This paper provides an in-depth analysis of the development and comparison of a physics-based model (PBM) and an equivalent-circuit model (ECM) for accurately capturing the dynamics of lithium-ion batteries across a broad operational range. The PBM integrates critical factors like hysteresis effects, concentration-dependent diffusivity, and the Arrhenius law for a realistic depiction of battery behaviors. The ECM is a second-order model with a detailed parameter map that accounts for SOC and temperature variations. To ensure a fair comparison, both the PBM and ECM are rigorously calibrated and validated through parameter identification and testing across 55 different operating conditions. To the best of the authors' knowledge, this represents the most comprehensive model calibration and validation effort for PBM and ECM in the literature to date, encompassing large temperature variations (-20 to 40°C), various charging/discharging C-rates, and real-world driving cycles. 
%Comparative analysis between the PBM and ECM highlights key differences in accuracy, computational complexity, parameterization requirements, and performance under varying temperature conditions. %The study underscores the potential of PBMs to outperform ECMs in accuracy and robustness, with comparable computational efficiency when challenges such as model order reduction and parameterization are effectively addressed. 
%This study offers valuable insights for the battery community in selecting appropriate models for battery management applications.
\end{abstract}


\end{frontmatter}

% \linenumbers

\doublespacing

\newpage
\section{Introduction}
Lithium-ion batteries are the preferred technology for many applications such as consumer electronics, electric vehicles, and energy storage systems, due to their high energy density, decreasing cost, and long service life \cite{masias2021opportunities, xu2023high}. 

Despite their widespread adoption, lithium-ion batteries face several challenges that hinder their broader application. An instance includes the influence of climatic variations on battery performance and the related battery management strategies. For example, approximately 38\% of regions in China may experience winter temperatures plummeting as low as -20°C, with certain areas witnessing substantial temperature differentials exceeding 40°C throughout the year \cite{chang2020aqueous}. These fluctuations lead to substantial changes in battery performance over a broad temperature range, presenting significant challenges in developing precise and robust control strategies within the battery management system (BMS). These strategies are essential for tasks such as state estimation, optimal charging, and fault diagnosis and prognosis.

Given that there are two main model categories used in BMS — equivalent circuit models (ECMs) \cite{dang2016open, zhu2019state} and physics-based models (PBMs) \cite{doyle1993modeling, chaturvedi2010algorithms} — it raises an important question about comparing their attributes: accuracy, robustness, adaptability, and computational complexity. To determine the superior choice, a thorough evaluation is needed to assess which model better meets the overall requirements.

PBMs are based on first principles such as thermodynamics, reaction kinetics, and mass transport and can capture key battery behaviors with high fidelity and accuracy \cite{doyle1993modeling, haran1998theoretical}. Those models preserve the physical meaning of the parameters and provides detailed information about internal electrochemical processes within the cell. However, PBMs typically use partial differential equations (PDEs) with many variables and parameters, which can lead to significant computational requirements \cite{moura2016battery, prada2012simplified}.

On the contrary, ECMs are typically characterized by lumped-parameter ordinary differential equation (ODE) systems, making them intuitive and simple to develop with low computational effort. However, because ECM parameters lack physical meaning, extensive testing and calibration are required to broaden their validity and maintain high accuracy under typical operating conditions \cite{xing2014state, lai2018comparative}.

In recent years, many model order reduction techniques have been developed to reduce the computational complexity of PBMs, such as Pad{\'e}  approximation \cite{marcicki2013design}, Galerkin projection \cite{fan2017reduced}, and polynomial approximation \cite{subramanian2005efficient}. While these techniques have notably enhanced computational efficiency, there has been no comprehensive study comparing the computation times of reduced-order PBMs and ECMs. Are PBMs still significantly lagging behind ECMs in computational efficiency, or have they already become nearly comparable? This is one of the key questions that this study aims to address.

Additionally, PBMs have a large number of model parameters, making precise parameterization much more challenging than for ECMs. It is well-known that nonlinear optimizations generally yield worse accuracy for problems with more parameters due to the increased likelihood of local minima in the cost function. To mitigate the risks of local minima and ensure robust model parameterization, various techniques have been implemented. These include the use of sensitivity analysis to reduce the number of parameters to be identified \cite{van2019stepwise, fan2021global}, the application of physical constraints on certain parameters (e.g., measurements of electrode thickness and particle size \cite{wassiliadis2022systematic}), careful selection of cost functions for optimization \cite{wang2024reduced}, and the employment of Pareto multi-objective optimization approaches \cite{zhang2014multi, wang2024reduced}. Therefore, it would also be interesting to examine the robustness of PBMs and ECMs under various operating conditions.

There are a few existing comparative studies between ECMs and PBMs in the literature \cite{meng2018overview, berrueta2018comprehensive, liang2022comparative}; however, almost all these studies only focus on a limited set of aspects, resulting in partial and incomplete analyses. For example, they often compare model performance under a narrow range of operating conditions. Additionally, a potential issue with these scenarios is that the model parameters for the ECMs or PBMs may not be fully optimized using data from such limited conditions. As a result, comparisons based on these models—if not fully optimized or validated—are likely to be unfair.

To address this gap, we will use 55 different operating conditions in this comprehensive comparative study of PBMs and ECMs. These conditions span a temperature range from -20 to 40$^{\circ}$C, include both charging and discharging at various C-rates, and incorporate real-world driving cycles based on field data. In addition, given that LFP/graphite batteries are rapidly gaining popularity in electric vehicles and stationary energy storage applications due to their low cost and balanced performance, they have been selected for this study. Additionally, other considerations include: 1) they exhibit a more pronounced hysteresis effect \cite{ovejas2019effects}, and 2) they demonstrate significant performance variations across a wide temperature range, particularly at low temperatures \cite{bae2014quantitative, tron2019aqueous}. Both of these factors present challenges that are crucial for developing accurate models.

Based on the above discussion, this paper makes the following three key contributions:

1) A comprehensive PBM is developed to accurately capture battery dynamics across a wide operational range, incorporating the hysteresis effect, concentration-dependent diffusivity, adjustments to Butler-Volmer kinetics, and the Arrhenius law.
2) Model parameter identification and subsequent validation are performed using 55 different operating conditions, covering both charging and discharging at various C-rates, and including real-world driving cycles based on field data. To the best of the authors' knowledge, this represents the most extensive model calibration and validation effort for PBM and ECM in the literature to date.
3) Based on the validated PBM and ECM, the models are compared across several key aspects, including accuracy, computational complexity, the number of parameters to be calibrated or updated, and performance across different temperature ranges. These comparisons are expected to offer insights into the strengths and limitations of each modeling approach, helping to inform and guide future developments and applications in battery modeling. 

The rest of the paper is structured as follows. The following section provides a brief introduction of the experimental data used in this study. Then, Section \ref{sec:model} presents the development of the PBM and ECM. Next, detailed results about the battery hysteresis characterization, model parameter identification and validation, as well as the multidimensional comparison of the PBM and ECM are presented in Section \ref{sec:results}. Summary and conclusions of this work are presented in the final section.

\section{Experimental Setup}
\label{sec:exp_setup}
The prismatic battery used in this study has a capacity of 166 Ah, with a LFP cathode and a graphite anode. Its operating voltage range is from 2.5V to 3.65V. The ambient temperature was controlled by a RH-GDW-100L environmental chamber, and current profiles at different charge/discharge conditions were applied with an Arbin LBT 5V-500A battery tester, sampling at a frequency of 1 Hz. 

\begin{figure}[!h]
\centering
\includegraphics[width=1\columnwidth]{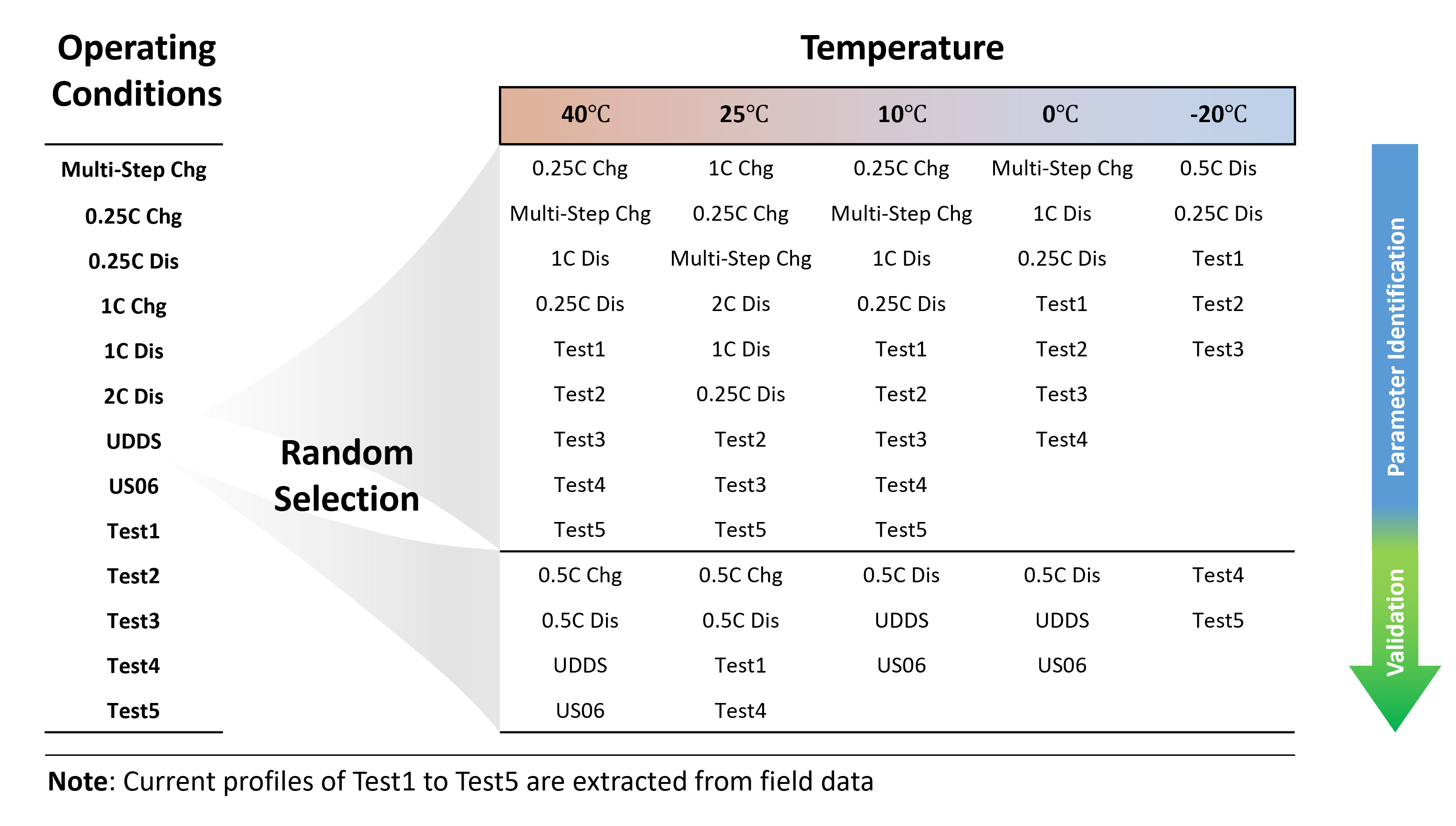}
\caption{Experimental data for model performance assessment.}
\label{fig:DOE}
\end{figure}
 
Figure \ref{fig:DOE} shows 55 experimental sets for characterizing and validating the PBM and ECM. All experiments were conducted in an environmental chamber and repeated at different temperatures (-20$^{\circ}$C, 0$^{\circ}$C, 10$^{\circ}$C, 25$^{\circ}$C, and 40$^{\circ}$C). At each temperature, the battery was charged and discharged at different C-rates (e.g., 0.25C, 1C, 2C). Additionally, a multi-step charging protocol was also used to investigate model performance in fast-charging scenarios. Dynamic driving cycle tests such as UDDS, US06, and Tests 1 to 5 (extracted from field data) were also considered to evaluate model performance in real-world applications. It is worth mentioning that to ensure each battery starts with the same capacity at different temperatures, all batteries were first fully charged at 25°C using a constant current constant voltage (CCCV) protocol, with the cutoff current set at 1/20C duirng the CV phase. They were then tested at different temperatures once thermal equilibrium had been reached.

Among these tests at various operating conditions, some data were randomly selected for model parameter identification, while the rest were used for subsequent model validation. At low temperatures such as -20$^{\circ}$C and 0$^{\circ}$C, the maximum C-rates in dynamic profiles were reduced according to battery specifications to avoid potential lithium plating with high currents. Similarly, the overall number of tests was also reduced accordingly.

\section{Model development}
\label{sec:model}

(1) \textbf{The reduced-order PBM}. The reduced-order PBM developed in this study starts from an extended single particle model, which is based on the porous electrode theory. This model integrates electrochemical kinetics to describe the internal electrochemical behavior of a lithium battery along the thickness of the electrode and separator, as well as the radius of the electrode particles. It has been widely demonstrated to achieve a good balance between maintaining accuracy and managing computational complexity \cite{fan2016modeling, fan2017reduced, fan2023nondestructive}.

In our previous work \cite{wang2024reduced}, we have developed a reduced-order physics-based battery model for NMC/graphite batteries for a wide temperature range, by considering 1) concentration-dependent diffusion, 2) adjustment of Butler-Volmer (B-V) equation, accounting for the excess driving force of Li$^+$ (de)intercalation in the charge transfer reaction in the presence of large concentration gradients \cite{xiong2023decoupled, xiong2023overpotential}, and 3) segmented Arrhenius equation. Please refer to \cite{wang2024reduced} for more details. In this work, in addition to these modifications, to address the strong hysteresis effect for LFP/graphite batteries, the Plett hysteresis model \cite{plett2015battery, gao2022enhanced} was incorporated into the physics-based battery model. 

\begin{figure}[ht]
\centering
\includegraphics[width=1\columnwidth]{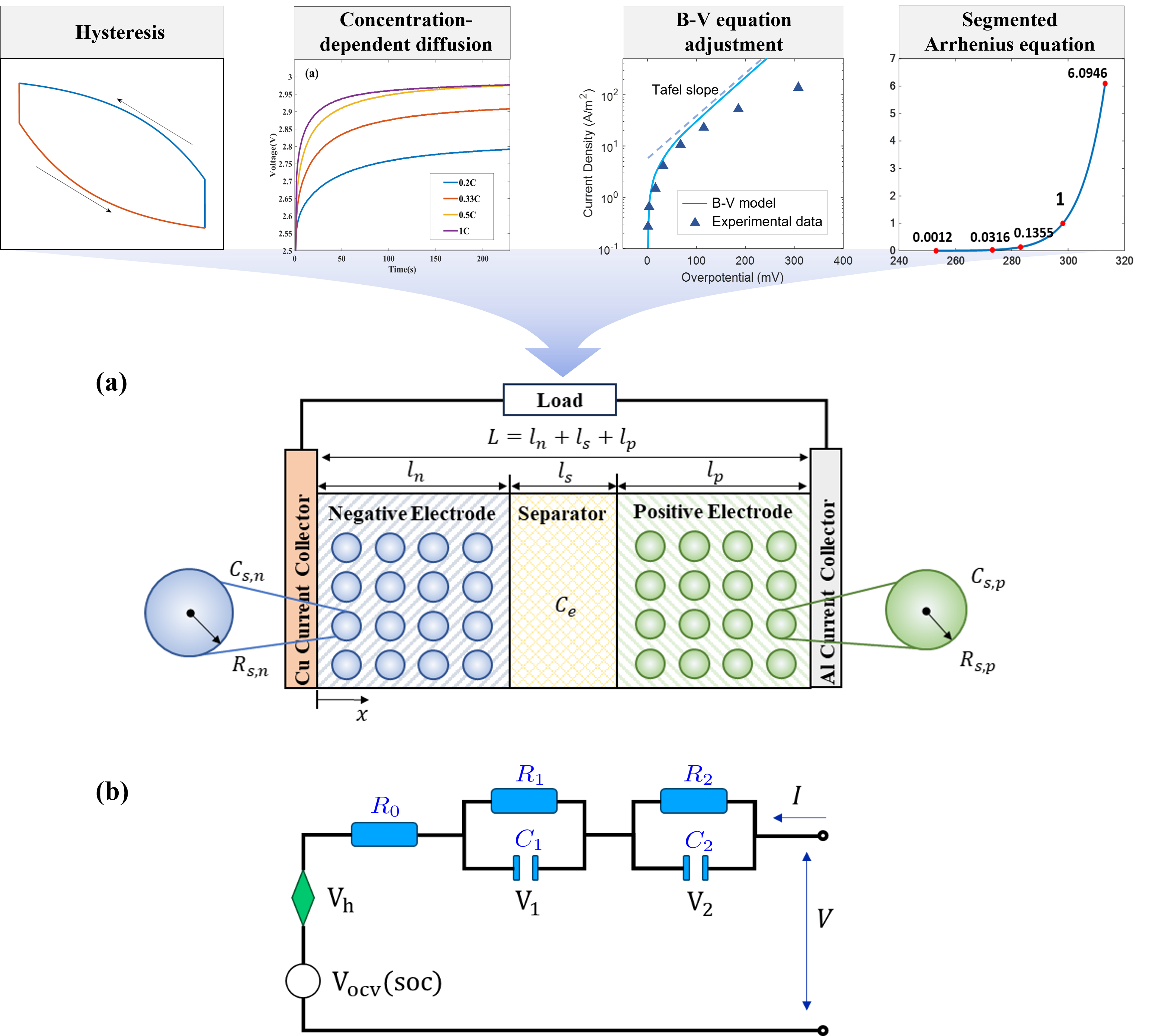}
\caption{The PBM and ECM used in this study. (a) Modified PBM developed for wide temperature range and various operating conditions. (b) Second-order ECM}
\label{fig:PBM_ECM}
\end{figure}

In the Plett model, the hysteresis voltage consists of an instantaneous hysteresis voltage and a dynamic SOC-varying hysteresis, and can be expressed as:
\begin{equation}
\label{eq:Vh}
V_h=M_0s[k]+Mh[k]
\end{equation}

The instantaneous hysteresis voltage $M_0s[k]$ varies when the sign of the current changes:
\begin{equation}
\label{eq:M0s}
s[k] = \begin{cases}  -\text{sgn}(i[k]),  & \mbox{$|i[k]|>0$}\\
  s[k-1], & \mbox{$i[k]=0$}
\end{cases}
\end{equation}

The dynamic hysteresis voltage $Mh[k]$ changes as the SOC changes, where $M$ is a function that gives the maximum polarization due to hysteresis as a function of SOC and the rate of change of SOC, and the hysteresis state $h$ is given by:
\begin{equation}
\label{eq:h}
h[k+1]=\exp\left(-\left|\frac{\eta[k]i[k]\gamma \delta t}{Q}\right|\right)h[k]-\left(1- \exp\left(-\left|\frac{\eta[k]i[k]\gamma \delta t}{Q}\right|\right)\right)\text{sgn}(i[k])
\end{equation}

%The three SOC-dependent hysteresis parameters, $\gamma$, $M_0$, and $M$ are then determined using Levenberg–Marquardt algorithm in MATLAB for lowest RMSE between the data and the model.

Figure \ref{fig:PBM_ECM}(a) illustrates the schematic of the modified physics-based battery model, which has been specially adapted for a wide temperature range and various conditions. The governing equations of the PBM used in this work are listed in Table \ref{tab:ESPM_Governing_Eqs}. Details about the model order reduction to the PDEs are skipped here and can be found in \cite{fan2017reduced}.

\newcounter{tblEqCounter} %create a counter
\setcounter{tblEqCounter}{\theequation}
\begin{table}[!h]
\fontsize{8pt}{4pt}\selectfont
\renewcommand{\arraystretch}{6.7}
\caption{Governing equations of the PBM used in this work.}
\label{tab:ESPM_Governing_Eqs}
\begin{center}%\hspace*{-2cm}
\addtolength{\leftskip} {-1.5cm}
\addtolength{\rightskip} {-1.5cm}
\begin{tabular}{lll}
\toprule
    &   Governing equations  &   \\
\midrule
Solid phase diffusion PDE  & $ \displaystyle \frac{\partial c_{s,k}}{\partial t} =  \displaystyle \frac{D_{s,k}}{r^2} \displaystyle \frac{\partial}{\partial r}\left( r^2  \displaystyle \frac{\partial c_{s,k}}{\partial r}\right),$  \quad $(k=p,n)$ &  \numberTblEq{eq:solid_PDE} \\
Boundary conditions & $\left. \displaystyle \frac{\partial c_{s,k}}{\partial r}\right|_{r=0} = 0$ &\\
 & $\left.D_{s,k} \displaystyle \frac{\partial c_{s,k}}{\partial r}\right|_{r=R_k} = -\displaystyle \frac{J_k\left(t\right)}{F} $ , \quad where $D_{s,k}=D_{s,k,\text{ref}}\exp(\mu_k \text{abs}(\Delta c^{\text{surf}}_{\text{bulk}}))$ \cite{wang2024reduced}  &  \multirow{-2}*{\numberTblEq{eq:solid_BCs}}  \\
Liquid phase diffusion  PDE & $\varepsilon_{e,k}\displaystyle \frac{\partial c_{e}}{\partial t} = \displaystyle \frac{\partial}{\partial x} \left(D_e \displaystyle \frac{\partial c_e}{\partial x}\right) + a_k \displaystyle \frac{1-t_0^+}{F}J_k,$ \quad where $(k=p,s,n)$ & \numberTblEq{eq:liquid_PDE} \\
Boundary conditions & $\left. \displaystyle \frac{\partial{c_e}}{\partial{x}}\right|_{x=0} = 0$, \quad    \quad   $\left. \displaystyle \frac{\partial{c_e}}{\partial{x}}\right|_{x=L} =0$  &\\
   & $\left.c_{e}\right|_{x=L_p^-} = \left.c_{e}\right|_{x=L_p^+}$  , \quad    \quad   $\left.c_{e}\right|_{x=L_p+L_s^-} = \left.c_{e}\right|_{x=L_p+L_s^+}$  &\\
   & $-D_{e,p}\left. \displaystyle \frac{\partial c_{e}}{\partial x}\right|_{x=L_p^-}=-D_{e,s}\left.\displaystyle \frac{\partial c_{e}}{\partial x}\right|_{x=L_p^+}$  , \quad    \quad 
 $-D_{e,s}\left.\displaystyle \frac{\partial c_{e}}{\partial x}\right|_{x=L_p+L_s^-}=-D_{e,n}\left. \displaystyle \frac{\partial c_{e}}{\partial x}\right|_{x=L_p+L_s^+}$  & \multirow{-3}*{\numberTblEq{eq:liquid_BCs}}\\
%\midrule
%& \multicolumn{2}{l}{Additional equations}\\
%\midrule
Intercalation current density &  $\displaystyle J_k = \begin{cases}  \displaystyle \frac{I}{a_k A L_k} & \mbox{, $k = p$}\\
\displaystyle 0 & \mbox{, $k = s$}\\
  -\displaystyle  \frac{I}{a_k A L_k} & \mbox{, $k = n$}
\end{cases}$   &   \multirow{-1}*{\numberTblEq{eq:J}} \\
Exchange current density    &  $i_{0,k}=Fk_{0,k}\sqrt{c_{s,k}^{\text{surf}}(c_{s,\max,k}-c_{s,k}^{\text{surf}})c_e}$   &   \multirow{-1}*{\numberTblEq{eq:exchange_current}} \\
Normalized Li$^+$ concentration & $\theta_k=\displaystyle \frac{c_{s,k}^{\text{surf}}}{c_{s,\text{max},k}} $   &  \numberTblEq{eq:normalized_theta} \\
Overpotential    &  $\eta_k=\eta_{\text{ct},k} + \eta_{\text{in},k} $ \cite{xiong2023decoupled}  &\\ 
& $\eta_{\text{ct},k}=\displaystyle \frac{\bar{R}T}{\alpha F}\sinh^{-1}\left(\frac{J_k}{2i_{0,k}}\right)$, \quad   $\eta_{\text{in},k} = \begin{cases}  \displaystyle \rho \frac{c_{s,\max,k}}{c_{s,k}^{\text{surf}}}J_k &  \mbox{, $0<\theta<\theta_c$}\\
\displaystyle \rho \frac{c_{s,\max,k}}{c_{s,\max,k}-c_{s,k}}J_k &  \mbox{, $\theta_c<\theta<1$} \end{cases}$ & \multirow{-2}*{\numberTblEq{eq:Bulter-Volmer2}} \\
Electrolyte potential   & $\displaystyle \frac{\partial{\phi_e}}{\partial{x}} =-\displaystyle \frac{i_e}{\kappa^{\text{eff}}}+ \displaystyle \frac{\kappa^{\text{eff}}_D}{\kappa^{\text{eff}}} \displaystyle \frac{\partial{\ln c_e}}{\partial{x}}$, \ \ \  where $\kappa^{\text{eff}}_D=\displaystyle \frac{2\bar{R}T\kappa^{\text{eff}}(1-t^{+}_0)}{F}(1+\beta)$   &   \multirow{-1}*{\numberTblEq{eq:PhiE_PDE_v2}} \\
Cell voltage with hysteresis   &  $V(t)=\phi_s(0,t) - \phi_s(L,t) -IR_c + V_h$ \cite{plett2015battery} & \\
& $ \quad \quad \ =  \big(U^p(\theta^p(t))-U^n(\theta^n(t))\big) $ &\\
& $  \ \ \ \ \ \ \ \ \ \ - \left(\displaystyle \frac{\bar{R}T}{\alpha F}\sinh^{-1}\left(\frac{I\left(t\right)}{2a_p A L_p i_{0,p}(t)}\right) -  \frac{\bar{R}T}{\alpha F}\sinh^{-1}\left(\frac{-I\left(t\right)}{2a_n A L_n i_{0,n}(t)}\right) \right) $ & \\
& $  \ \ \ \ \ \ \ \ \ \ - \displaystyle \frac{I(t)}{2A}\left(\frac{L_p}{\kappa^{\text{eff}}} + \frac{2L_s}{\kappa^{\text{eff}}} + \frac{L_n}{\kappa^{\text{eff}}} \right) + \frac{\kappa^{\text{eff}}_D}{\kappa^{\text{eff}}} \left[ \ln c_e(0,t) - \ln c_e(L,t) \right] -IR_c + M_0s[k] +Mh[k]  $ &   \multirow{-5}*{\numberTblEq{eq:ESPM_voltage}} \\
\bottomrule
\multicolumn{3}{l}{\footnotesize Subscriptions: $p$ = positive electrode, $s$ = separator, $n$ = negative electrode.}\\
\end{tabular}
\end{center}\hspace*{-2.5cm}
\end{table}
\setcounter{equation}{\thetblEqCounter} %at the end of the table, set the equation numbering to the counter

(2) \textbf{The second-order ECM with hysteresis}. In this study, the second-order ECM in this work comprises a typical equivalent circuit network with two RC pairs. Additionally, the ECM model also integrates the previously introduced Plett model (Eq.~\ref{eq:Vh}) to consider the hysteresis effect of LFP/graphite batteries, as shown in Fig.~\ref{fig:PBM_ECM}(b).

%\begin{figure}[!h]
%\centering
%\includegraphics[width=0.55\columnwidth]{Figures/ECM.png}
%\caption{The schematic of a second-order ECM.}
%\label{fig:ECM}
%\end{figure}

The voltages of the RC pairs, SOC and the terminal voltage $V$ are given as: 

\begin{equation}
\begin{aligned}
\label{eq:V_ECM}
& V_1[k+1] = V_1[k]\exp\left(-\frac{\Delta t}{R_1C_1} \right) + I[k] R_1\left( 1-\exp\left(-\frac{\Delta t}{R_1C_1} \right)\right) \\
& V_2[k+1] = V_2[k]\exp\left(-\frac{\Delta t}{R_2C_2} \right) + I[k] R_2\left( 1-\exp\left(-\frac{\Delta t}{R_2C_2} \right)\right) \\
& SOC[k+1]=SOC[k] + \frac{-\eta[k] \Delta t}{Q\cdot 3600}\cdot I[k] \\
& V[k]=V_{\text{OCV}}(SOC[k])-I[k]  R_{0}-V_1[k]-V_2[k] + V_h[k]
\end{aligned}
\end{equation}
where the parameters $R_0$, $R_1$, $R_2$, $C_1$ and $C_2$ are functions of both SOC and temperature, and need to be identified.

\section{Results and Discussion}
\label{sec:results}
\subsection{Hysteresis characterization}
\begin{figure}[!h]
\centering
\begin{subfigure}{0.85\textwidth}
  \centering
  \includegraphics[width=1\columnwidth]{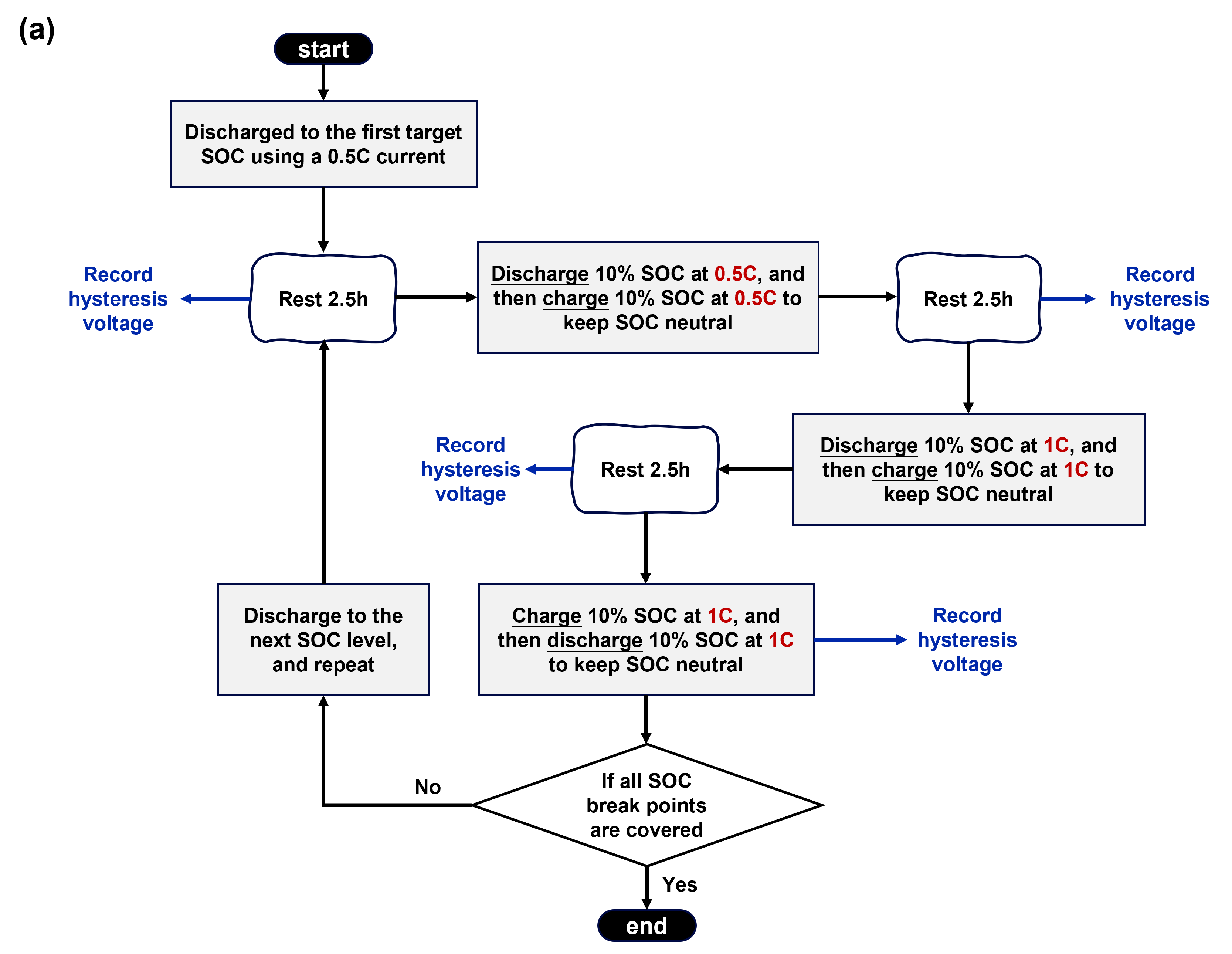}
\end{subfigure}
%\vspace{3ex}

\begin{subfigure}{1\textwidth}
  \centering
  \includegraphics[width=1\columnwidth]{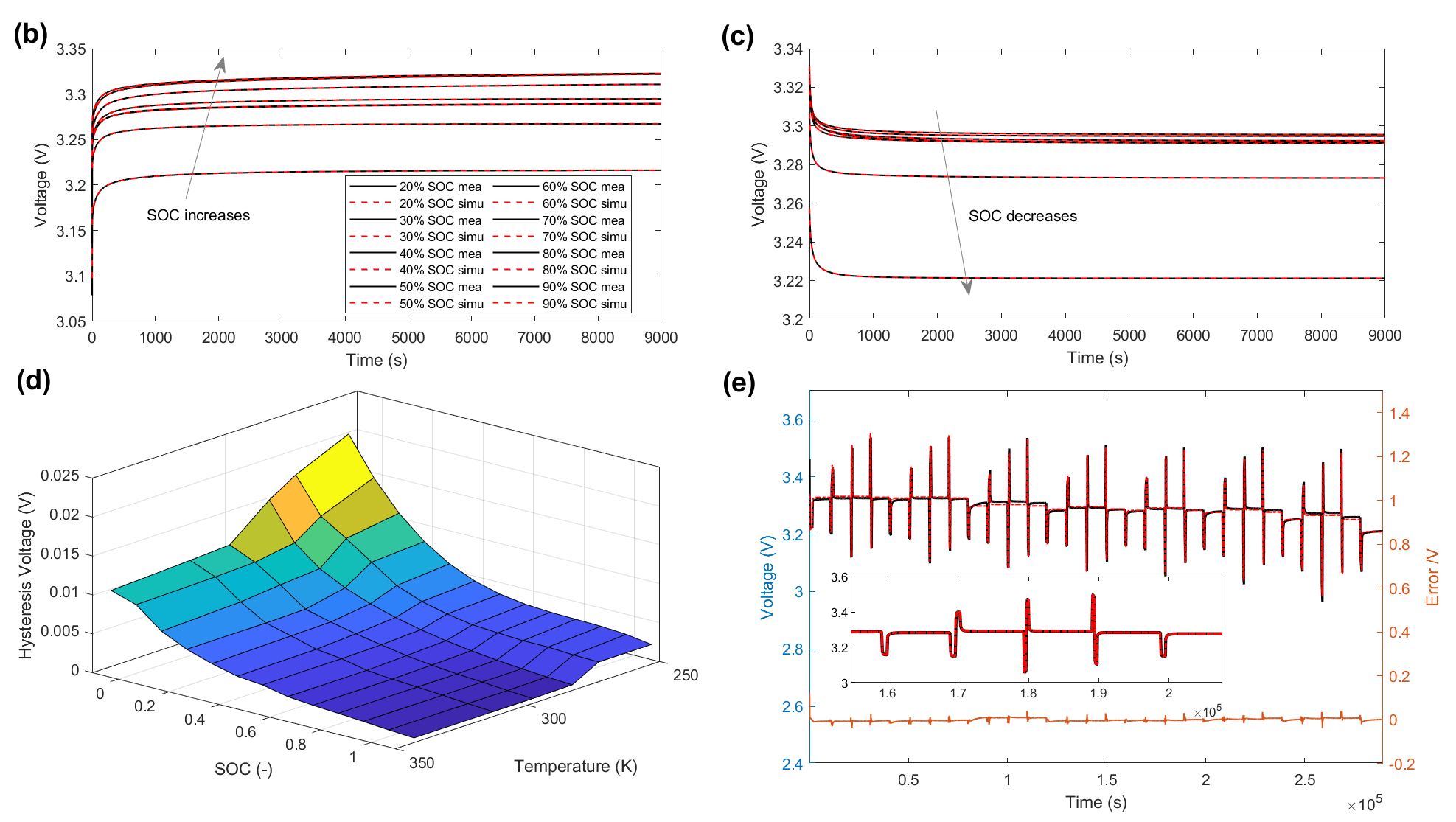}
\end{subfigure}
\caption{Characterization of battery hysteresis effect. (a) Hysteresis characterization test procedure. (b) Prediction of OCV after discharge and hours of relaxation. (c) Prediction of OCV after charge and hours of relaxation. (d) Hysteresis map as a function of SOC and temperature. (d) Validation of the Plett hysteresis model.}
\label{fig:hysteresis}
\end{figure}

Although the hysteresis is not thoroughly understood in literature yet, it is widely recognized that it is influenced by SOC, temperature and even the direction of current. Thus, to accurately capture the battery hysteresis voltage, a comprehensive hysteresis characterization test was conducted at each temperature. The test procedure is outlined in Fig.~\ref{fig:hysteresis}(a). 

The test profile is designed to explore the parameter space (e.g., SOC and current direction) and to induce cell hysteresis effects under these conditions. At each temperature, the process begins with the cell being fully charged to 100\% SOC. It is then discharged to the first target SOC using a 0.5 C current, followed by a 2.5-hour rest period. A series of constant current pulses of 0.5 C and 1 C is then applied. Each charge pulse is applied after a corresponding discharge pulse to maintain a neutral SOC. The duration of each pulse is set to induce a 10\% change in SOC starting from the target SOC. Next, an additional set of 1 C current pulses is applied, with the charge pulse preceding the discharge pulse, to account for hysteresis with respect to current direction. After each pulse set, there is a 2.5-hour rest period to record the hysteresis voltage. The cell is then either discharged to the next SOC level using a 0.5 C current. 

It is noted from Fig.~\ref{fig:hysteresis}(b) and (c) that at the end of the rest periods, the voltages at some SOC points do not reach a full equilibrium state and continue to evolve slowly after 2.5 hours. To obtain accurate hysteresis voltages, we employed a two-step approach: 1) fitting the 2.5-hour voltage data using a simple exponential function with respect to time, $k_1t^{k_2}+k_3$, where $k_1$, $k_2$, and $k_3$ are the fitting coefficients, and 2) predicting the equilibrium voltage or OCV after 8 hours of relaxation. This method proved to be effective at all SOC points during the rest periods following the discharge/charge pulses, as indicated by the red dashed lines in Fig.~\ref{fig:hysteresis}(b) and (c). The same test repeats at several times at -20, 0, 10, 25 and 45$^{\circ}$C, and the full hysteresis voltage map is then obtained as a function of SOC and temperature, as shown in Fig.~\ref{fig:hysteresis}(d).

%An example of the voltage profile of the characterization test at 25$^{\circ}$C is shown in Fig.~\ref{fig:hysteresis}(b).

%Next, the parameters in the Plett hysteresis model, $M_0$, $M$ and $\gamma$ are obtained using . 
Figure \ref{fig:hysteresis}(e) shows an example of the voltage profile of the characterization test at 25$^{\circ}$C. To obtain the parameters in the Plett hysteresis model, $M_0$, $M$ and $\gamma$, the particle swarm optimization (PSO) algorithm \cite{fan2020systematic} was employed to minimize the differences between the model prediction values and experimental data. As shown in comparison results, the Plett model with identified parameter agrees very well with experimental data across the whole SOC range. The root mean square error in voltage is only 6.7mV. The results demonstrate that the Plett model with identified parameters accurately captures the battery's hysteresis behavior across various SOC levels. %The identified values of $M_0$, $M$ and $\gamma$ are summarized in Table XXX.

%\begin{figure}[!h]
%\centering
%\includegraphics[width=1\columnwidth]{Figures/Hysteresis.eps}
%\caption{Characterization of battery hysteresis effect. (a) Modified PBM developed for wide temperature range and various operating conditions. (b) Second-order ECM}
%\label{fig:PBM_ECM}
%\end{figure}

\subsection{Model parameter identification and validation}
The rest of model parameters in the PBM and ECM were then obtained using the experimental data that introduced in Section \ref{sec:exp_setup} and the PSO algorithm. A summary of the comparison setup is outlined in Table~\ref{fig:comparison}.

\begin{table}[h]
  \caption{Comparison setup for the PBM and ECM.}
  \label{fig:comparison}
  \includegraphics[width=1\columnwidth]{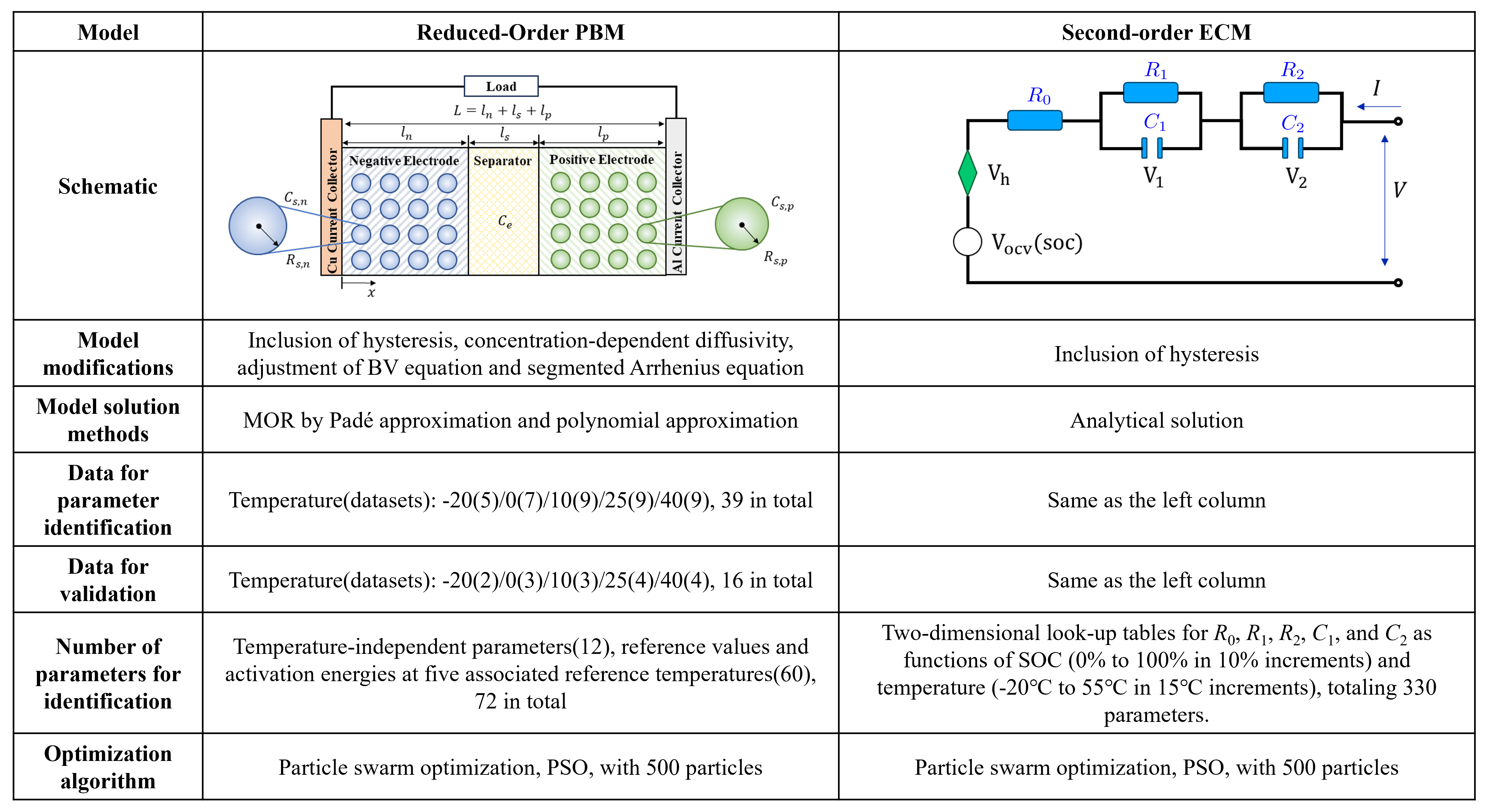}
\end{table}

\textbf{Model solution methods}: The solid-phase PDEs in the PBM were approximated and solved using the Pad{\'e}  approximation, while the liquid-phase PDEs were addressed with polynomial approximation. Those model order reduction (MOR) techniques have been proven to significantly reduce the computational complexity of the PDEs in the PBMs with merely marginal loss of accuracy \cite{marcicki2013design, fan2017reduced, li2021reduced}. On the other hand, the second-order ECM was solved analytically in a state-space representation by capturing the dynamics of the RC pairs \cite{plett2015battery}.

\textbf{Experimental data assignment}: Experimental data were divided into two groups for parameter identification (or model calibration) and validation. For parameter identification, 39 datasets were selected, covering various charging/discharging conditions across a wide temperature range of -20 to 40$^{\circ}$C. The remaining 16 datasets were used to validate the model performance with the identified parameter values. To ensure a fair comparison, the datasets in each group for the PBM and ECM were the same.

%In the PBM, there are 12 temperature-independent parameters such as the electrode surface area, thickness of the electrodes and separator. Temperature-dependent parameters include the diffusion coefficients, kinetics rate constants and ionic conductivity. As mentioned previously, a single set of parameters (i.e., the activation energies) of the Arrhenius equation may no longer work well for a wide temperature range \cite{liang2022comparative}. For example, the negative solid phase diffusion coefficient reduces significantly at temperatures below zero, and its decrease across the entire temperature range does not follow an exact exponential function. To address this issue, a segmented Arrhenius modeling approach is used to address the drastic property changes in those parameters due to the wide temperature range \cite{wang2024reduced}. The wide temperature range of -20 to 40$^{\circ}$C is divided into four segments at -3, 7 and 20$^{\circ}$C. Within each temperature range, the pattern of parameter variation still follows the Arrhenius equation, with the previous segment's parameter value at the given segmentation point as the reference value. The segmented Arrhenius modeling approach results in 60 parameters, and 72 to be identified in total in the entire PBM.

\textbf{Number of parameters for identification}: In the PBM, there are 12 temperature-independent parameters, such as the electrode surface area and the thickness of the electrodes and separator. Temperature-dependent parameters include diffusion coefficients, kinetics rate constants, and ionic conductivity. It has been demonstrated that a single set of parameters (e.g., the activation energies) for the Arrhenius equation may not perform well across a wide temperature range \cite{kohout2021modified, liang2022comparative}. For example, the negative solid-phase diffusion coefficient significantly decreases at temperatures below zero, and its variation across the entire temperature range does not follow an exact exponential function. To address this, a segmented Arrhenius modeling approach \cite{wang2024reduced} is used in this work to capture the drastic property changes of these parameters over the wide temperature range. The temperature range of -20 to 40$^{\circ}$C is divided into several segments at -17, -5, 10, 30 and 38$^{\circ}$C. Within each segment, parameter variation follows the Arrhenius equation. This segmented approach results in 60 additional parameters (reference values and associated activation energies at 5 segmented temperature points), with a total of 72 parameters to be identified in the entire PBM.

	As for the second-order ECM, two-dimensional parameter look-up tables are typically established for $R_0$, $R_1$, $R_2$, $C_1$ and $C_2$, as functions of SOC and temperature \cite{miniguano2019general, kwak2019parameter}. Overall, there are 330 parameters (0\% to 100\% in 10\% increments in SOC and -20 to 55$^{\circ}$C in 15$^{\circ}$C increments in temperature) in the ECM. Note that the higher temperature limit of 55$^{\circ}$C is used because cell temperatures can exceed the ambient temperature of 40$^{\circ}$C during charging and discharging. 

\textbf{Optimization algorithm}: The PSO algorithm is applied in this study due to its easy implementation, fast convergence, and strong ability to reduce risks of local minima. For both the PBM and ECM parameter identification tasks, 500 particles were selected. The cost function for parameter identification is defined as:
\begin{equation}
\underset{\hat{\boldsymbol{\Theta}}}{\text{minimize}} \sum_{i=1}^{M}\sqrt{ \frac{ \displaystyle \sum_{t=1}^{t_f} \left( {V}_i(t) - \hat{{V}}_i(t,\hat{\boldsymbol{\Theta}})\right)^2}{N_i}}
\label{eq:cost_fun}
\end{equation}
where $M$ is the number of input test datasets, $t_f$ is the end of time for each test, ${V}$ and $\hat{{V}}$ are the measured and predicted voltages, $N_i$ is the number of data points, and $\hat{\boldsymbol{\Theta}}$ is the parameter set to be identified. The PSO algorithm computations are executed in Matlab using parallel computing on a cluster equipped with dual Intel Xeon ICX Platinum 8358 CPUs (64 cores) and 512 GB of RAM.

\subsection{Multidimensional comparison of the PBM and ECM}

\begin{figure}[!h]
\centering
\includegraphics[trim={70 60 70 60},clip,width=1\columnwidth]{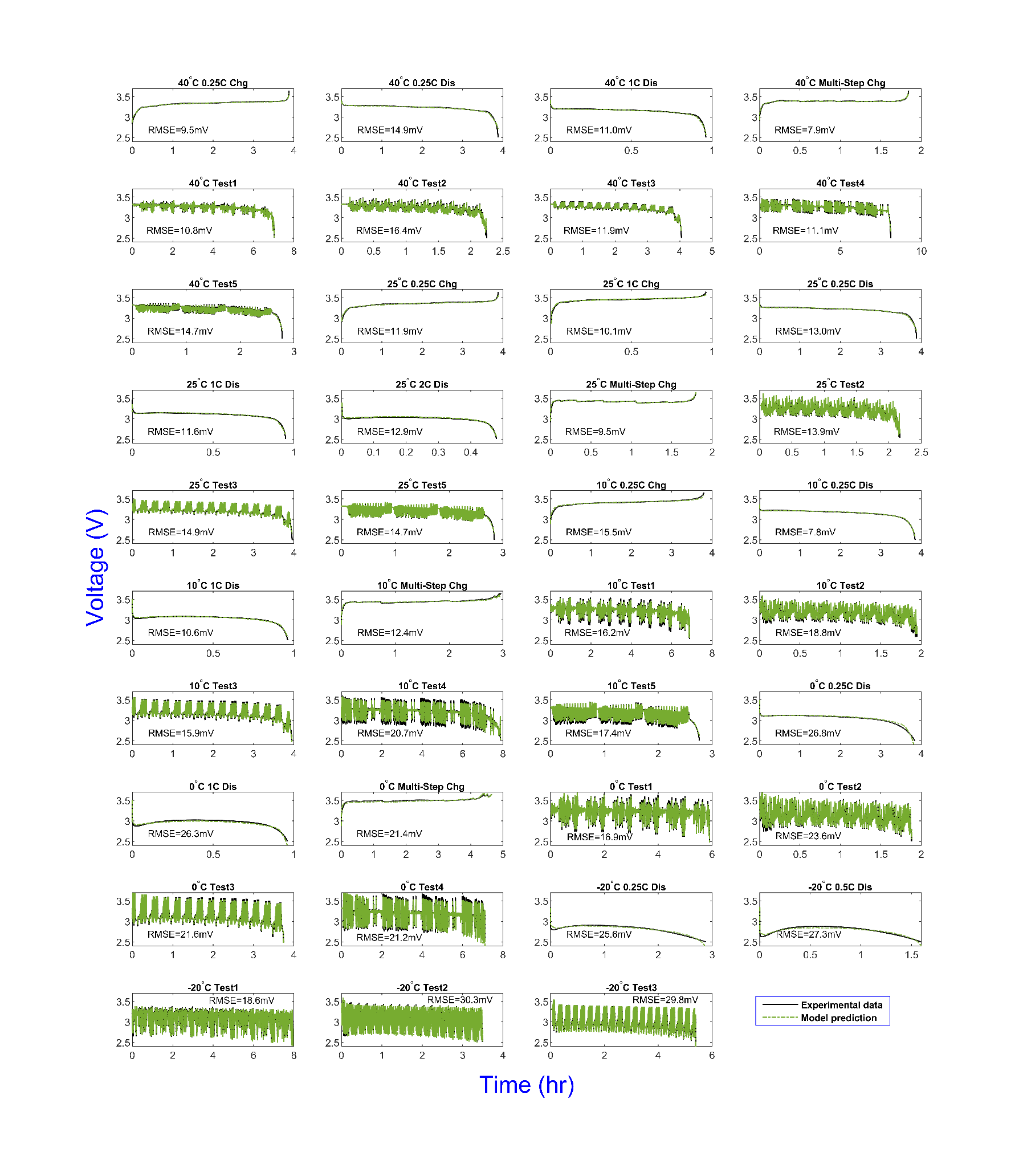}
\caption{Model parameter identification results for the PBM.}
\label{fig:param_id_PBM}
\end{figure}

\begin{figure}[!h]
\centering
\includegraphics[trim={70 60 70 60},clip,width=1\columnwidth]{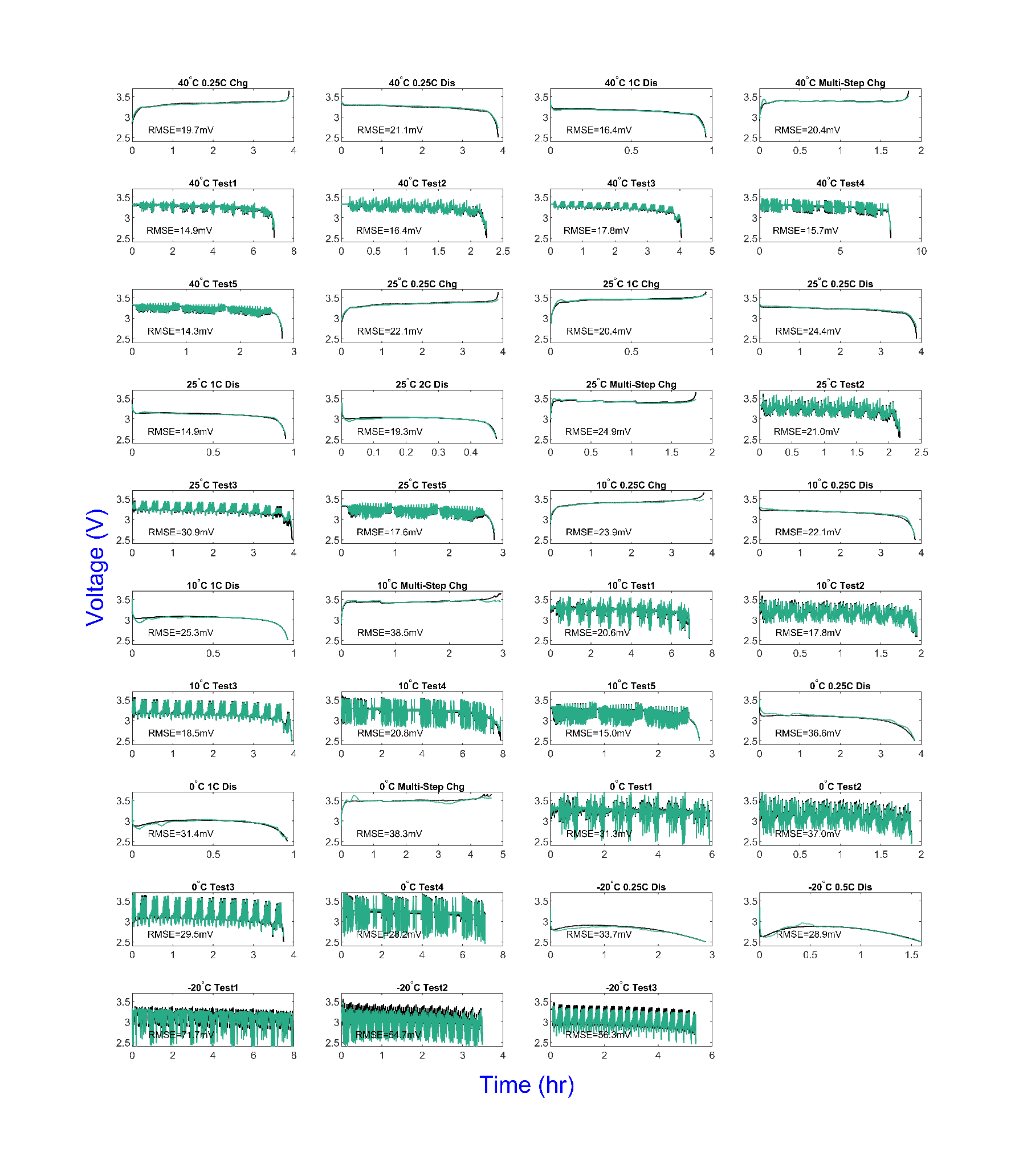}
\caption{Model parameter identification results for the ECM.}
\label{fig:param_id_ECM}
\end{figure}

Figures~\ref{fig:param_id_PBM} and \ref{fig:param_id_ECM} present the model parameter identification results for the PBM and ECM under 39 operating conditions. Each subplot also includes the RMSE between the model predictions and experimental data. Generally, both the PBM and ECM with optimized parameters show better agreement with experimental data at higher temperatures, likely due to reduced kinetic and concentration overpotentials from higher ionic conductivity and diffusivity. The optimized models accurately capture battery behaviors during constant rate charging, discharging, multi-step charging, and various dynamic driving cycles.

However, when temperatures drop below 0$^{\circ}$C, the nonlinear dynamics of the batteries increase significantly, leading to larger nonlinear overpotentials. For instance, the discharge voltage curves at 0.25C and 0.5C at -20$^{\circ}$C exhibit "arch" shapes, distinctly different from those at higher temperatures. This might be due to a slight increase in internal battery temperature during operation, reducing overall overpotentials slightly. Yet, given the relatively low C-rate (e.g., 0.25C) and limited temperature rise (less than 5$^{\circ}$C increase during the test), the "arch" shape of the voltage response is likely due to significant kinetic overpotential at high SOC. Despite optimized model parameters, it remains challenging for the models to maintain the same level of accuracy as at higher temperatures, resulting in a considerable increase in prediction errors.

%\begin{figure}[!h]
%\centering
%\includegraphics[width=1\columnwidth]{Figures/Fig3.png}
%\caption{Experimental data for model performance assessment.}
%\label{fig:comparison}
%\end{figure}

%It is also worth noting that even if the ECM achieves overall good performance for these 39 conditions, it shows increasing local prediction errors at certain SOC regions under some conditions, such as low SOC during the multi-step charging at 40$^{\circ}$C and high SOCs during the 1C discharging at 0$^{\circ}$C and 0.5C discharging at -20$^{\circ}$C in Fig.~\ref{fig:param_id_ECM}. As these SOC- and temperature-dependent parameters of the ECM do not have physical meanings and are empirically fitted to minimize the overall errors of all conditions, it is sometimes inevitable that the fitted parameters work well for some SOCs or temperatures, but may not for others. This is one of the major limitations of the ECMs that it requires significant data to ensure its validity range. On the other hand, as shown in Fig.~\ref{fig:param_id_PBM}, the high-fidelity PBM show consistent performance at different temperatures, SOCs across different input profiles, demonstrating great potential to accurately capture the physical variations of internal battery states.
It is important to note that while the ECM performs well overall across these 39 conditions, it exhibits increasing local prediction errors in certain SOC regions under specific conditions. For instance, unexpected errors are observed at low SOC locally during multi-step charging at 40$^{\circ}$C and at high SOC during 1C discharging at 0$^{\circ}$C and 0.5C discharging at -20$^{\circ}$C, as shown in Fig.~\ref{fig:param_id_ECM}. Since the SOC- and temperature-dependent parameters of the ECM are empirically fitted without inherent physical meanings, they are primarily aimed at minimizing overall errors across all conditions. However, this approach can result in good performance for some SOCs or temperatures, while failing for others. This is a major limitation of ECMs: their effectiveness is relatively limited across a broad range of operating conditions. In contrast, as demonstrated in Fig.~\ref{fig:param_id_PBM}, the high-fidelity PBM shows consistent performance across various temperatures and SOCs, accurately capturing the physical variations of internal battery states across different input profiles.

\begin{figure}[!h]
\centering
\includegraphics[trim={70 10 70 10},clip,width=1\columnwidth]{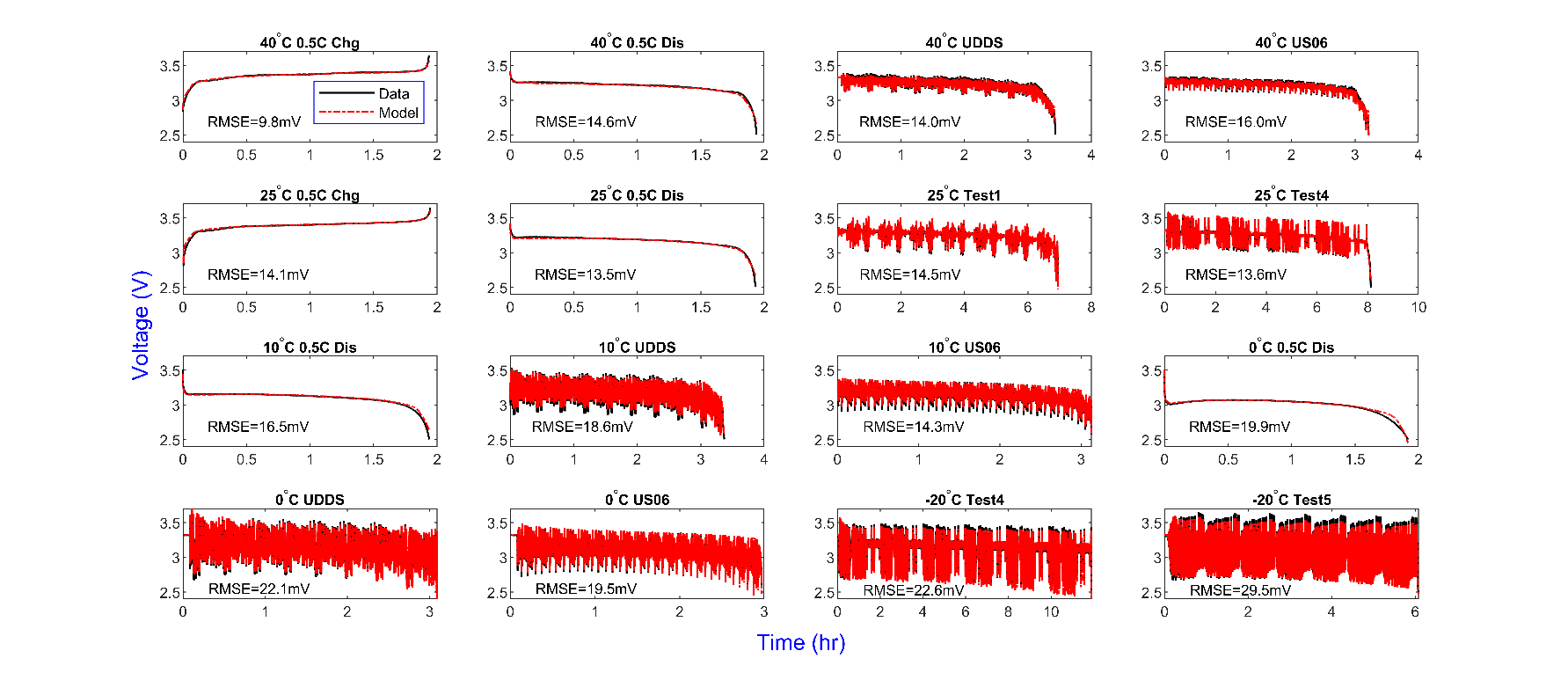}
\caption{Model validation results for the PBM.}
\label{fig:validation_PBM}
\end{figure}

\begin{figure}[!h]
\centering
\includegraphics[trim={70 10 70 10},clip,width=1\columnwidth]{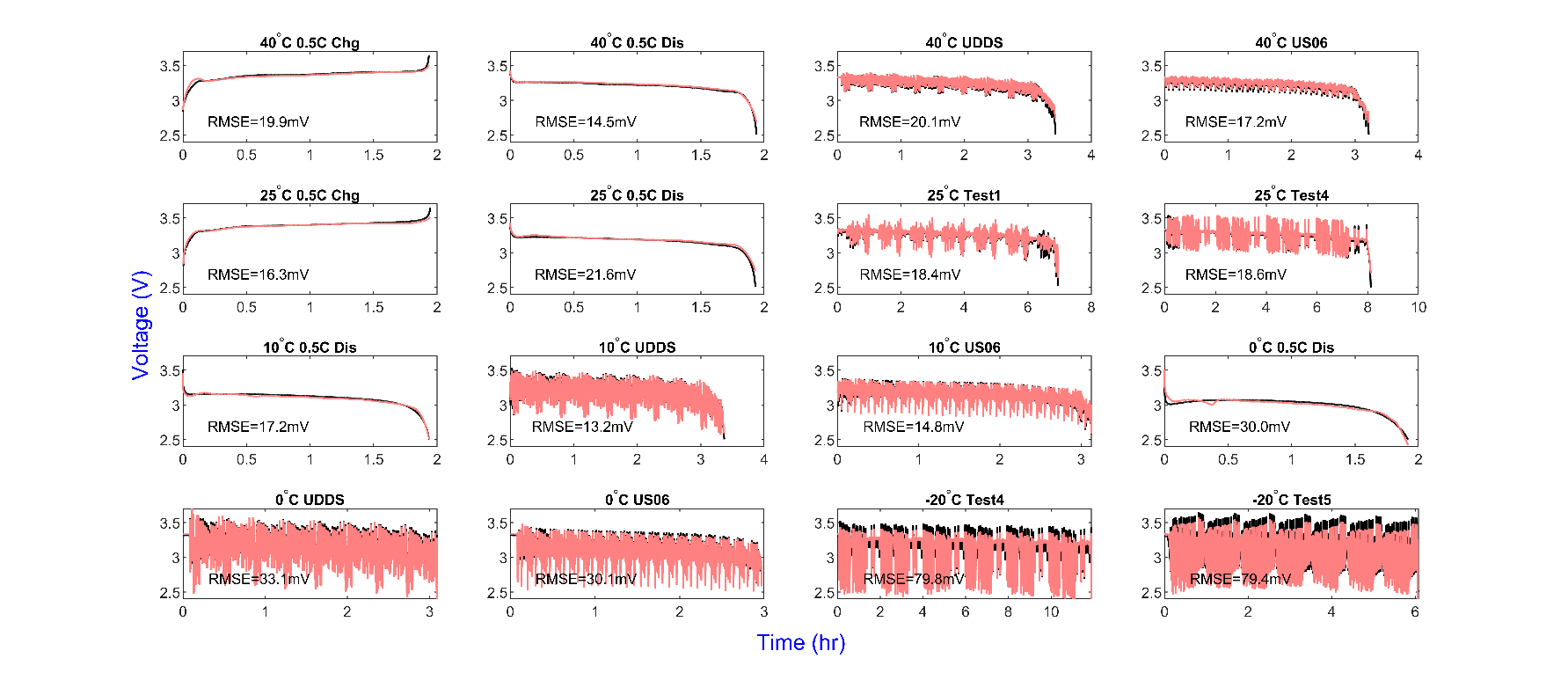}
\caption{Model validation results for the ECM.}
\label{fig:validation_ECM}
\end{figure}

%With model parameters optimized by the 39 datasets across different operating conditions and temperatures, the PBM and ECM with identified model parameters are then further validated against data from 16 other operating conditions. To this end, similar observations can be found in Figs.~\ref{fig:validation_PBM} and \ref{fig:validation_ECM}. Both the PBM and ECM perform better at higher temperatures than low temperatures. Nevertheless, by carefully comparing the individual RMSEs, it is clear that the PBM can make more accurate predictions than the ECM.
With model parameters optimized using 39 datasets across various operating conditions and temperatures, the PBM and ECM were further validated against data from 16 additional operating conditions. As shown in Figs.~\ref{fig:validation_PBM} and \ref{fig:validation_ECM}, both models perform better at higher temperatures compared to lower temperatures. However, a careful comparison of the individual RMSEs reveals that the PBM consistently provides more accurate predictions than the ECM.

It is also worth mentioning that all RMSEs in the above figures were all calculated over the entire SOC range, from 2.5V to 3.6V. Accurately capturing voltage responses at low SOC ranges below 20\% is particularly challenging due to large variations in the open-circuit potentials at the electrodes and various sources of overpotentials. Nonetheless, the PBM still aligns very well within these SOC ranges across different operating conditions. A detailed summary of the comparison between the PBM and ECM is also provided in Table~\ref{fig:table_comparison}, highlighting the accuracy of both models across all temperatures and operating conditions during the parameter identification and validation phases.

\begin{table}[!h]
  \caption{Summary of the comparison of the PBM and ECM at the parameter identification and validation phases.}
  \label{fig:table_comparison}
  \includegraphics[width=1\columnwidth]{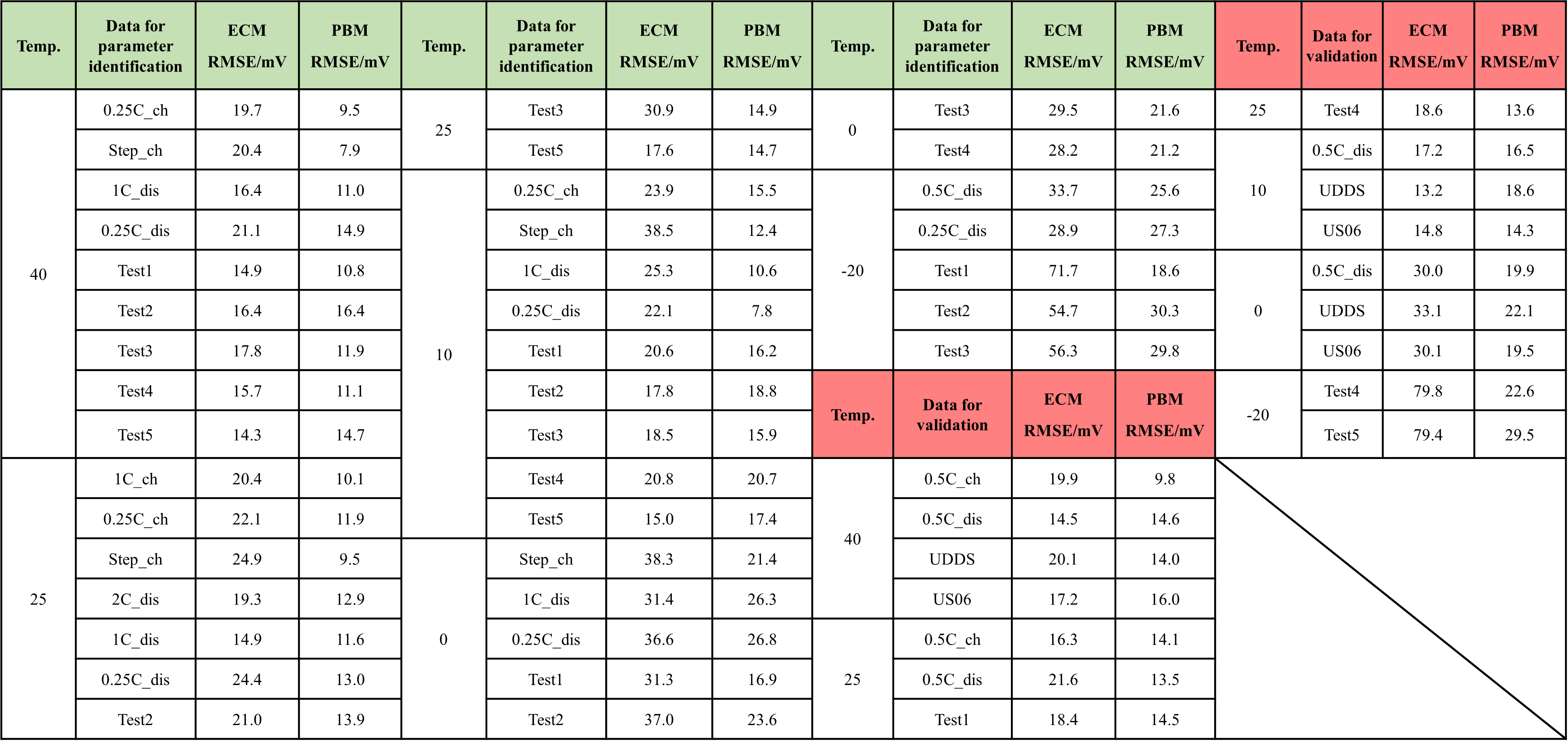}
\end{table}

%Figure~\ref{fig:rmse_comparison} compares the performance of the PBM and ECM using the averag RMSE at each temperature. While the observations are consistent with those in Figs.~\ref{fig:param_id_PBM}-{fig:validation_ECM}, it's shown that the error of the ECM at low temperatures starts to increase dramatically at the validation case, while there is only a mild error increase by the PBM from 0 to -20$^{\circ}$C. This is a good example showing the predictive capabilities of the PBM, deriving from first principles. On the other hand, extensive testing for calibration are required to broaden their validity range and to maintain high level of accuracy at various operating conditions.

%Figure~\ref{fig:rmse_comparison} compares the performance of the PBM and ECM using the average RMSE at each temperature. The observations are consistent with those in Figs.~\ref{fig:param_id_PBM} to \ref{fig:validation_ECM}. It is shown that the error of the ECM at low temperatures increases dramatically in the validation case, while the PBM only exhibits a mild error increase from 0 to -20$^{\circ}$C. This illustrates the predictive capabilities of the PBM, by accurately capturing the internal processes such as thermodynamics, mass transport and charge transfer kinetics. On the other hand, extensive testing and calibration are required to broaden the validity range and maintain a high level of accuracy under various operating conditions.

\begin{figure}[!h]
\centering
\includegraphics[width=0.8\columnwidth]{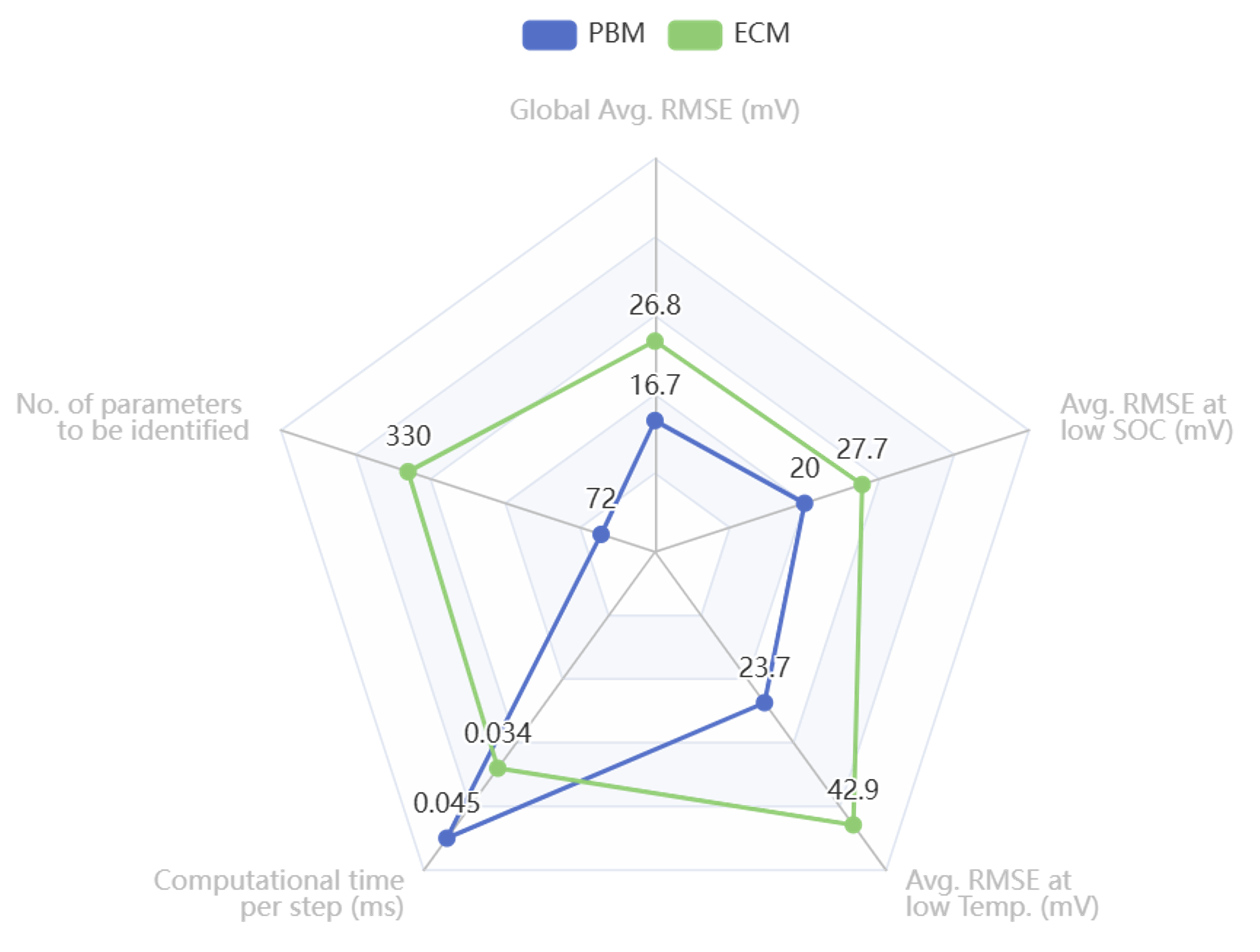}
\caption{Model validation results for the ECM.}
\label{fig:spider_comparison}
\end{figure}

Figure~\ref{fig:spider_comparison} presents a comprehensive comparison between the PBM and ECM, focusing on the overall average RMSE, accuracy at low SOC (below 20\%) and low temperatures (0 and -20$^{\circ}$C), computational time per sampling step, and the total number of tunable parameters. The results demonstrate that the PBM outperforms the ECM, particularly with significantly smaller errors at low SOC and temperature ranges, leading to better overall accuracy across all operating conditions. Although the computational time for the PBM is slightly higher than that of the ECM, they are nearly comparable, at 0.045ms and 0.034ms, respectively. Furthermore, the PBM requires only 72 tunable parameters for wide operating applications, in contrast to the 330 parameters needed by the ECM. This reduction in the number of parameters, combined with higher accuracy, is achieved by precisely incorporating first-principle processes within the batteries.

%The discussion above highlights that developing an accurate PBM requires extensive expertise to thoroughly characterize the physical and chemical properties of battery components, along with deep knowledge of the mathematical modeling of internal physical and electrochemical processes. Additionally, to extend the application of PBMs to online scenarios, efficient model order reduction techniques are essential to minimize computational costs while preserving accuracy. Equally important is the careful parameterization of the model to ensure its accuracy and robustness across various operating conditions. Once these challenges are addressed, PBMs can offer superior performance in terms of accuracy, robustness, and computational efficiency.
%
%In contrast, the ECM is simpler to develop and implement due to its straightforward model structure. However, its accuracy and robustness can significantly diminish under challenging conditions, such as at low SOCs and temperature ranges, where nonlinear battery behaviors become more pronounced. Special attention is required when applying ECMs under these conditions.

Based on the above discussion, it is evident that developing an accurate PBM necessitates extensive expertise in both the characterization of the physical and chemical properties of battery components and the mathematical modeling of the internal physical and electrochemical processes occurring within the battery. A deep understanding of these complex interactions is essential to create a model that accurately reflects the real-world behavior of the battery. Furthermore, to extend the applicability of PBMs to broader contexts, such as online monitoring and control scenarios, it is crucial to implement efficient model order reduction techniques. These techniques are necessary to significantly lower the computational cost associated with PBMs, without compromising the model’s accuracy. In addition, the process of model parameterization must be meticulously executed to ensure that the PBM remains accurate and robust across a wide range of operating conditions. Proper parameterization is key to maintaining the model’s reliability, especially when dealing with the variable and often harsh conditions encountered in practical applications. Once these challenges are effectively addressed, PBMs have the potential to outperform other modeling approaches in terms of accuracy, robustness, and computational efficiency, making them highly valuable for advanced BMS.

On the other hand, the ECM offers simplicity in development and implementation due to its straightforward model structure. However, this simplicity comes with trade-offs. The accuracy and robustness of ECMs can deteriorate significantly under challenging conditions, such as low SOC and extreme temperatures, where the battery's nonlinear behaviors become more pronounced. In these scenarios, the limitations of the ECM can lead to noticeable errors. Therefore, special attention must be given when employing ECMs in these conditions to avoid compromising the reliability of the model.

\section{Conclusions} \label{Section:conclusions}

In this study, we developed a comprehensive PBM that accurately captures the complex dynamics of battery behavior across a wide operational range. This model incorporates critical factors such as the hysteresis effect, concentration-dependent diffusivity, adjustments to Butler-Volmer kinetics, and the Arrhenius law, ensuring a detailed and realistic representation of battery performance. To validate the model, we conducted an extensive parameter identification process and subsequent validation using data from 55 different operating conditions. These conditions span both charging and discharging processes at various C-rates and include real-world driving cycles based on field data. This effort represents, to the best of our knowledge, the most comprehensive model calibration and validation for PBM and ECM presented in the literature to date. We then compared the validated PBM and ECM across several critical dimensions, including accuracy, computational complexity, the number of parameters requiring calibration or updating, and performance across different temperature ranges. These comparisons provide valuable insights into the strengths and limitations of each modeling approach, guiding future applications and development in battery modeling.

Developing an accurate PBM requires deep expertise in characterizing battery components and modeling internal physical and electrochemical processes. In addition, efficient model order reduction and rigorous parameterization procedures are essential to extend PBM applications while maintaining accuracy and robustness. When these challenges are addressed, PBMs can excel in accuracy, robustness, and computational efficiency, making them ideal for advanced BMS. In contrast, the ECM is simpler to develop and implement but may suffer from reduced accuracy and robustness under challenging conditions, such as low SOC and extreme temperatures. Caution is needed when using ECMs in these scenarios to ensure reliable performance.

%Future work will expand the study to incorporate aging models, both PBMs and ECMs, and evaluate the effectiveness of all three modeling approaches in predicting capacity fade.

%Future work will investigate 
%
%kinetic parameters

\section*{Acknowledgments}

This work is funded by the National Natural Science Foundation of China (Grant No. 52307246 and 52177218) and the Natural Science Foundation of Shanghai (Grant No. 23ZR1429100).

\hspace*{\fill}

\hspace*{\fill}

\bibliography{myreference}

\end{document}